\documentclass[twocolumn]{aastex631}
\usepackage{makecell}
\usepackage{amsmath}

\begin{document}

\title{Hunting for the First Explosions at the High-Redshift Frontier}

\author[0000-0002-6038-5016]{Junehyoung Jeon}
\affiliation{Department of Astronomy, University of Texas, Austin, TX 78712, USA}
\affiliation{Cosmic Frontier Center, The University of Texas at Austin, Austin, TX 78712, USA}
\author[0000-0003-0212-2979]{Volker Bromm}
\affiliation{Department of Astronomy, University of Texas, Austin, TX 78712, USA}
\affiliation{Cosmic Frontier Center, The University of Texas at Austin, Austin, TX 78712, USA}
\affiliation{Weinberg Institute for Theoretical Physics, University of Texas, Austin, TX 78712, USA}
\author[0000-0003-2237-0777]{Alessandra Venditti}
\affiliation{Department of Astronomy, University of Texas, Austin, TX 78712, USA}
\affiliation{Cosmic Frontier Center, The University of Texas at Austin, Austin, TX 78712, USA}
\author[0000-0001-8519-1130]{Steven L.~Finkelstein}
\affiliation{Department of Astronomy, University of Texas, Austin, TX 78712, USA}
\affiliation{Cosmic Frontier Center, The University of Texas at Austin, Austin, TX 78712, USA}
\author[0000-0003-4512-8705]{Tiger Yu-Yang Hsiao}
\affiliation{Department of Astronomy, University of Texas, Austin, TX 78712, USA}
\affiliation{Cosmic Frontier Center, The University of Texas at Austin, Austin, TX 78712, USA}

\email{junehyoungjeon@utexas.edu}

\begin{abstract}
The James Webb Space Telescope (JWST) has spectroscopically confirmed galaxies up to $z\sim14$, 300 Myr after the Big Bang, and several candidates have been discovered at $z\sim15-25$, with one candidate as high as $z\sim30$, only 100 Myr after the Big Bang. Such objects are unexpected, since theoretical studies have not predicted the existence of detectable galaxies at $z\sim30$. While any $z\sim30$ candidates may be contaminants at lower redshifts, we explore whether such extreme redshift sources could be consistent with hyper-energetic transient events linked to the formation of the first, metal-free, stars. Specifically, we consider pair-instability supernovae (PISNe), a predicted class of extreme thermonuclear explosions that leave no remnant behind. Using cosmological simulations, we investigate an overdense cosmic region, where star formation and subsequent PISNe occur at $z\sim30-40$, even within standard cosmology. Assessing the likelihood of such a region, the corresponding number of PISNe at $z\gtrsim20$, and their observed flux, we find that JWST has a non-negligible chance to detect a PISN event at extremely high redshifts. If a transient event were confirmed at $z\sim30$, this would provide a direct glimpse into the epoch of first star formation, dramatically extending the empirical reach of astronomy.

\end{abstract}


\section{Introduction} \label{sec:intro}

The frontier of observational cosmology has been pushed to increasingly high redshifts by the James Webb Space Telescope (JWST), with spectroscopically confirmed galaxies out to $z\gtrsim14$ \citep{Naidu2025_z14,Carniani2024}. Recently, photometrically selected candidates have been identified at even earlier times, to $z\sim25$, via the photometric dropout technique \citep{Perez2025,Castellano2025}, and even $z\sim32$ by spectral energy distribution (SED) fitting \citep{Gandolfi2025}, corresponding to only $\sim100$~Myr after the Big Bang. If confirmed, these indications of stellar activity in the extremely early Universe would challenge our understanding of first star and galaxy formation \citep[e.g.,][]{Bromm2011,Dayal2018,Somerville2025,Yung2025}. 

The suggested period for these ultra-high-redshift candidates coincides with the formation of the first stars in the Universe. Primordial gas from the Big Bang, composed of hydrogen and helium, will collapse to form metal-free, Population~III (Pop~III) stars at $z\sim20 - 30$ \citep{Bromm2013,Klessen2023,Bromm2002,Abel2002,Nakamura2001}. Pop~III stars are predicted to be more massive than the metal-enriched stars at subsequent epochs, reaching masses up to a few 100 M$_{\odot}$ \citep{Hosokawa2011,Hirano2014,Susa2013,Stacy2016,Hosokawa2016,Sugimura2020,Latif2022}. However, the inferred total stellar masses of Pop~III systems formed within minihalos are low ($\lesssim10^3$ M$_\odot$), and they are thus not expected to be directly observable at $z\gtrsim 20$ \citep{Schauer2023}. Alternatively, Pop~III stars may be detectable when they end their lives as hyper-energetic pair-instability supernovae (PISNe). Such an event occurs when a progenitor star with a mass between 140 and 260~M$_\odot$ undergoes an extreme thermonuclear explosion\footnote{For rapidly rotating progenitors, the PISN mass range may extend to lower masses, impacting event rates and observational characteristics \citep[e.g.,][]{Chatzopoulos2012,Smidt2015}.}, triggered by electron-positron production in its core \citep{Heger2003}. These reactions cause a rapid loss of radiation pressure, resulting in runaway collapse and the ignition of explosive oxygen and silicon burning that completely disrupts the star, with no remnant being left behind \citep{Barkat1967,Rakavy1967,Fraley1968,Bond1984,Fryer2001,Chen2014}, reaching absolute UV magnitudes up to $-22$, or even brighter values close to the initial peak \citep[e.g.,][]{Kasen2011,Dessart2013,Gilmer2017}. Motivated by the newly discovered ultra-high-redshift candidates, we explore the possibility of JWST detecting such transient phenomena triggered by the first stars at $z\sim30$, following up on earlier studies \citep[e.g.,][]{Hummel2012,deSouza2013,Whalen2013,Hegde2023,Hegde2025}. Recently, \citet{Ferrara2026} have also argued that some high-redshift candidates could be PISN events.

Because of their extreme explosion energies \citep[$\sim10^{53}$~erg for a $\sim250$ M$_\odot$ progenitor;][]{Heger2002}, PISNe may be observable even up to $z\sim30$ (see Section~\ref{sec:discussion}). Previous works have argued that PISNe, should they exist, could be observed even up to $z\sim25$ \citep{Weinmann2005}. The main challenge in observing these events is their small number density: As the predicted visibility time of individual PISN events is short, of order $\sim10$~yr in the observed frame \citep{Hummel2012}, and given the limited JWST survey area, previous studies found that the probability of detecting a PISN event at $z\gtrsim8$ is low, the expected number of events being $\lesssim0.1$ \citep{Hummel2012,Gabrielli2024,Venditti2024}. Lower redshifts have been proposed to be more promising for PISN detection with JWST, or with upcoming wide-field Roman Space Telescope surveys \citep{Moriya2022,Moriya2022_2,Regos2020}. The discovery of $z\sim30$ candidates prompts us to revisit this question, using cosmological simulations of highly-biased, overdense regions to trace the formation of the first stars in such extreme environments, and to examine the probability of detecting the resulting PISNe at the (high-$z$) tail-end of early star formation. 

We specifically consider these related questions: Does JWST observe a sufficiently large volume, including archival searches,  to capture possible host systems for a PISN at extremely high redshifts (Section~\ref{sec:obsprob})? Does JWST cover sufficiently long periods to detect a PISN event (Section~\ref{sec:eventprob})? Are PISN events luminous enough for JWST to observe them at such early times (Section~\ref{sec:brightness})?


\section{Methodology} \label{sec:methods}

We run cosmological simulations of biased high-density regions that are capable of producing multiple Pop~III stars and the resulting PISNe, employing the \textsc{gizmo} code \citep{Hopkins2015} that inherits the gravity solver from the \textsc{gadget-2} framework \citep{Springel2005}, and includes accurate Lagrangian hydrodynamics, here in the meshless finite-mass (MFM) implementation. We use a modified version of \textsc{gizmo} \citep{Liu2020,Jeon2023,Jeon2024} with updated sub-grid models for star formation and feedback (Section~\ref{star_formation_model}), primordial chemistry, cooling, and metal enrichment, which have been tested against high-resolution simulations of first galaxy formation and high-redshift observations \citep{Jaacks2018,Jaacks2019}.

\subsection{Initializing a Biased Region}\label{sec:biased}

To be able to form a sufficient number of Pop~III stars and their resulting PISNe as early as $z\sim30$, we need an extreme overdensity from primordial fluctuations. Zoom-in simulations of biased regions from larger parent simulations can be run in order to capture such an overdensity \citep[e.g.,][]{Bahe2017,Jeon2024}. However, to sample the rarest and most extreme regions, very large parent simulations are required, which can be computationally costly. 

An alternative method to simulate an overdense region is to artificially boost the amplitude of primordial density fluctuations \citep[e.g.,][]{Greif2011}. Then, when sampling randomly from the boosted fluctuations, the probability of choosing a biased, overdense region is much higher. Following this approach, we use standard \textit{Planck} cosmological parameters \citep{Planck2016}: $\Omega_m = 0.315$, $\Omega_b = 0.048$, $n_s = 0.966$, and $h = 0.6774$, but we change the value of $\sigma_8$ here. Instead of the $\sigma_8= 0.829$ observed by \textit{Planck}, we increase it to $\sigma_8=1.5$ This overdense region, while rare, is predicted to be observable in the survey volumes JWST has achieved so far (Section~\ref{sec:obsprob}). The initial conditions are generated with the \textsc{MUSIC} code \citep{Hahn2011} at $z=99$, including both gas (baryonic) and dark matter (DM) components. We choose a box with (comoving) side length 6 $h^{-1}$ cMpc, which translates to $\sim2-3$ arcmin at $z\sim30$, corresponding to the field of view of one JWST NIRCam pointing. The box contains $2\times512^3$ particles in total, including both gas and DM. The DM particle mass is $1.76\times10^5$ M$_\odot$, and that of the gas particles $3.16\times10^4$ M$_\odot$. The star-particle mass, which represents a stellar population, is $1.98\times10^3$ M$_\odot$. The simulation uses an adaptive gravitational softening length with a minimum value of $0.5$ $h^{-1}$ ckpc.

\begin{figure*}
\gridline{
\fig{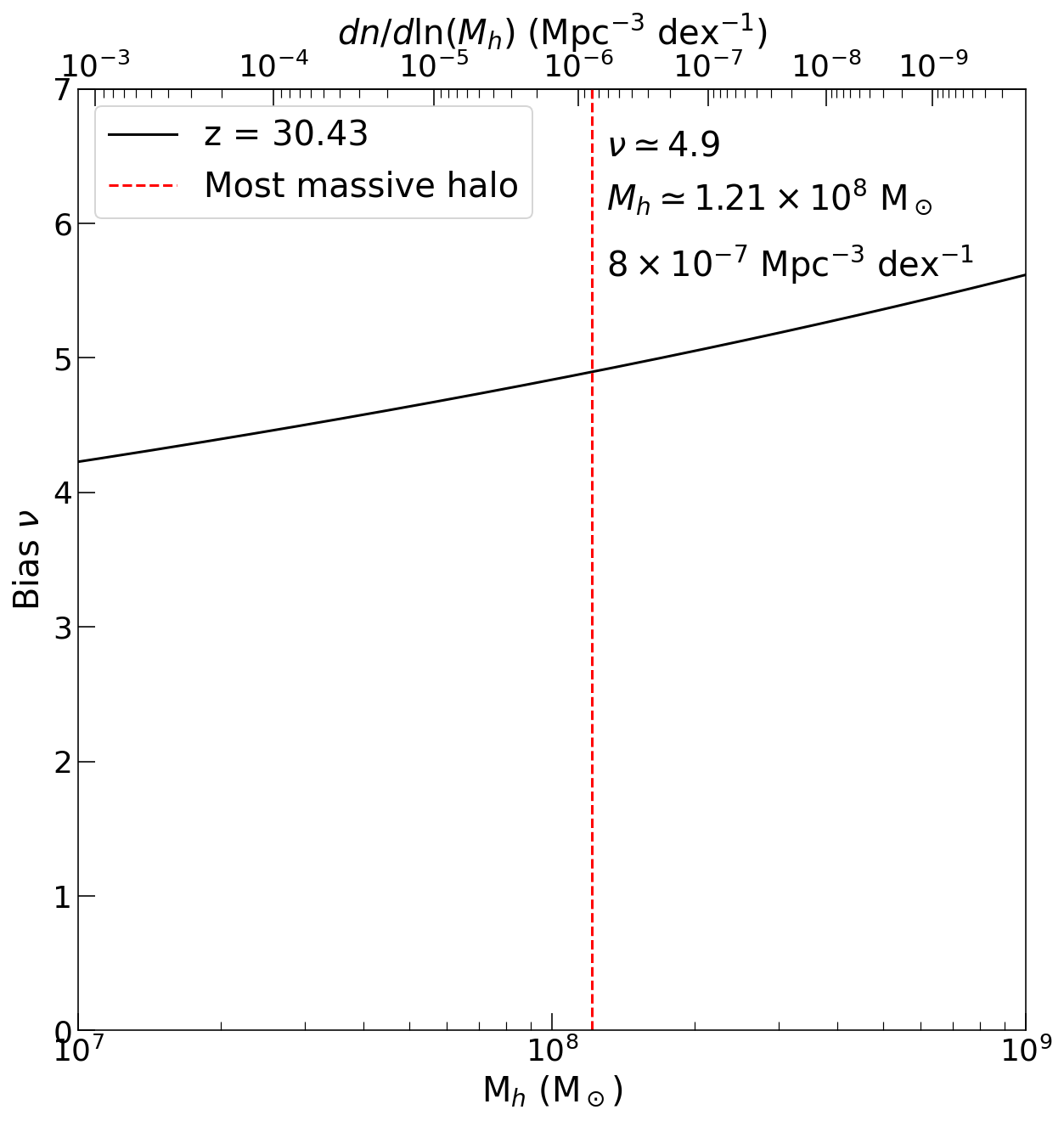}{0.45\textwidth}{}
\fig{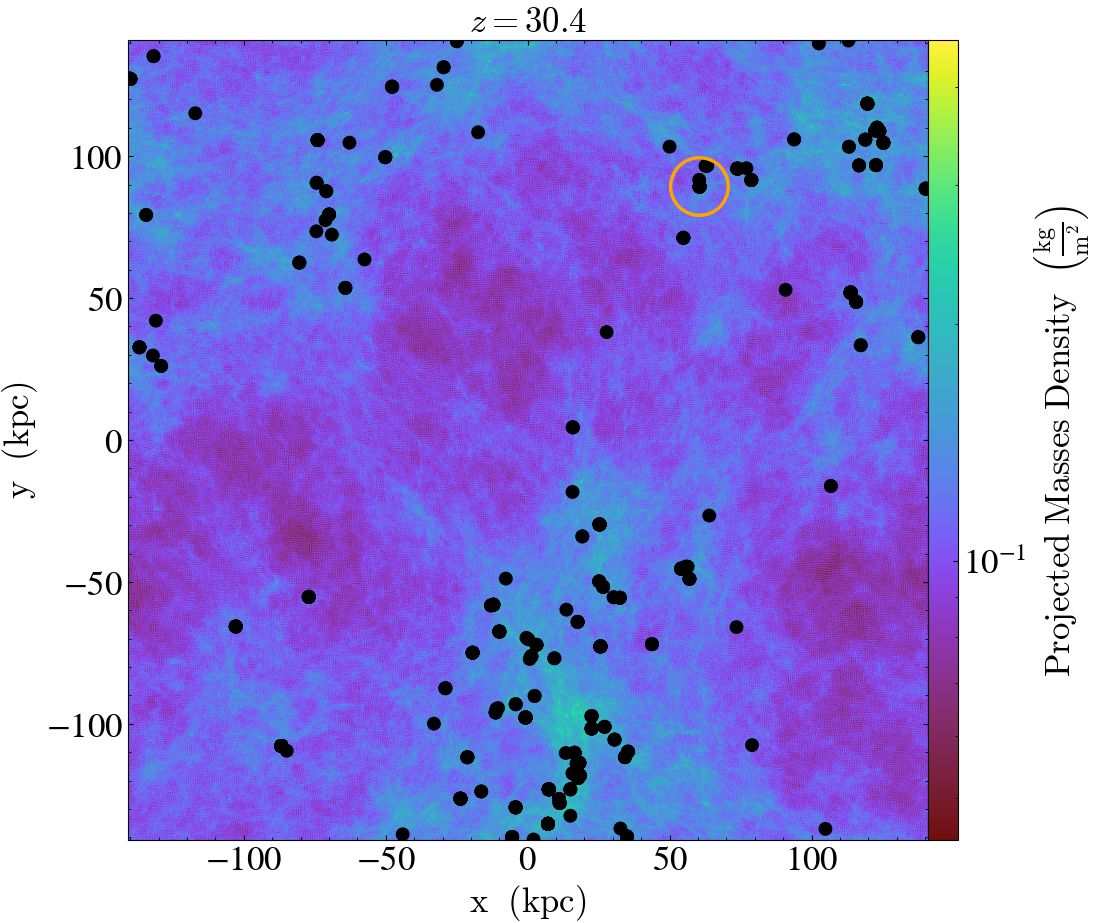}{0.5\textwidth}{}
}
    \caption{Biased, overdense simulation volume. \textit{Left:} The most massive dark matter halo in our simulation with $M_h\simeq1.2\times10^8$ M$_\odot$ at $z=30.4$ and the corresponding peak height $\nu\simeq5$. The corresponding number densities are also shown, giving $8\times10^{-7}$ Mpc$^{-3}$ dex$^{-1}$ for the most massive halo, based on the \textsc{GUREFT} halo mass function \citep{Yung2024}. A similar number density of $\sim10^{-6}$ Mpc$^{-3}$ dex$^{-1}$ is derived when using the \citet{Sheth1999} halo mass function. The overdense region is unlikely but not impossible to encounter in the early Universe. \textit{Right:} The projected gas density distribution of our simulation volume at $z\sim30.4$. The length is in physical units. Stellar particles are marked as black dots and the most massive halo as an orange circle, with sizes not to scale, showing that SF and the subsequent transient events can occur in such highly biased regions within the early Universe.} 
    \label{fig:siminit}
\end{figure*}

To assess the probability of realizing such a biased region within realistic cosmological density fluctuations, described by the standard $\sigma_8= 0.829$ parameter, we characterize the resulting bias by the peak height $(\nu)$ of the most massive halo in the simulation volume, defined as 
\begin{equation}
    \nu(M_h,z) = \frac{\delta_c}{\sigma(M_h,z)}\mbox{\ ,}
\end{equation}
where $\delta_c=1.686$ is the critical overdensity for spherical collapse, and $\sigma(M_h,z)$ the variance of the power spectrum on the scale of the halo with mass $M_h$ at redshift $z$. If $\nu<1$ at a given redshift, a halo of mass $M_h$ on average would have already collapsed at that redshift and if $\nu>1$, the halo on average will collapse in the future. We identify the dark matter halos in post processing using \textsc{Rockstar} \citep{Behroozi2013}, and the $\nu$ required for a given $M_h$ at $z=30.4$ is derived with the \textsc{Colossus} package \citep{Diemer2018} assuming the virial halo mass and radius definitions \citep{Bryan1998}. Fig.~\ref{fig:siminit} shows the most massive halo with mass $1.2\times10^8 ~\mathrm{M_\odot}$ and the corresponding peak height $\nu\sim5$ at $z=30.4$. We quantify the number density of halos of this mass using the halo mass function from the Gadget at Ultrahigh Redshift with Extra-Fine Timesteps (\textsc{GUREFT}) simulations, designed to capture the halo merger histories at extremely high-$z$ \citep{Yung2024}. The derived halo number density is $8\times10^{-7}$ Mpc$^{-3}$ dex$^{-1}$, or, employing the \citet{Sheth1999} halo mass function instead, $1.2\times10^{-6}$ Mpc$^{-3}$ dex$^{-1}$. Thus, the number density of the halo considered here, within the biased region, is approximately $\sim10^{-6}$ Mpc$^{-3}$ dex$^{-1}$, which renders our simulated volume extremely rare, but still achievable in the early Universe \citep[see also][]{Naoz2006}. We examine the likelihood of JWST observing such a region in Section~\ref{sec:obsprob}. 


\subsection{Star formation and feedback}\label{star_formation_model}

We adopt the stochastic star formation (SF) models developed and validated in previous studies \citep[see][]{Jaacks2018,Jaacks2019,Liu2020}. Specifically, the metal-free Pop~III models are calibrated to be consistent with extremely high-resolution zoom-in simulations, and the metal-enriched, Population~II (Pop~II) stellar feedback models to reproduce the observed star formation histories at $z\sim5-10$.

In the stochastic model, a gas particle becomes a SF candidate, when its hydrogen number density is $n_{\rm H}>100$ cm$^{-3}$ and its temperature $T\leq10^3$~K. For a SF candidate, the probability of spawning a stellar particle is 
\begin{equation}
    p_{\rm SF} = \frac{m_{\rm SF}}{m_*}[1-\exp(-\eta_*\Delta t/t_{{\rm ff},i})]\mbox{\ ,}
\end{equation}
where $m_{\rm SF}$ is the mass of the candidate gas particle, $m_*$ the mass of the stellar particle to be formed, $\eta_*$ the SF efficiency, $\Delta t$ the simulation timestep, and $t_{{\rm ff},i}= \sqrt{3\pi/(32G\rho_i)}$ the free-fall timescale of the gas particle with density $\rho_i$. 
For Pop~III stars, we set $\eta_*=0.05$ and for Pop~II $\eta_* = 0.1$ \citep{Jaacks2019}, reflecting the lower star-formation efficiencies of Pop~III minihalos. For both populations, we set $m_*\simeq600$ M$_\odot$ based on high-resolution Pop~III star formation simulations \citep{Bromm2013,Stacy2016,Klessen2023}.
A random number $p$ is generated for the uniform distribution $[0,1]$ and a stellar particle is formed when $p<p_{\rm SF}$. Fig.~\ref{fig:siminit} shows the resulting gas density distribution with the locations of stellar particles,demonstrating that in the overdense region simulated here SF and transients can occur as early as $z\sim30$. 


Since we here focus on tracking the locations and timing of the initial runaway collapse of the primordial gas, the subsequent stellar feedback is not 
relevant for the discussion below, regarding the detectability of transient events at extremely high redshifts, but the full details of the feedback physics can be found in \citet{Liu2020}. We do not directly measure the PISN event rate from the simulations, but instead derive it in post processing (Section~\ref{sec:eventprob}). Because we cannot resolve individual explosions at the simulation resolution \citep[e.g.,][]{Greif2007,Ritter2016}, we infer the PISN rate from the simulated SF rate density, assuming the Pop~III initial mass function (IMF) given below.

\section{Results} \label{sec:results}

\subsection{Observing the Biased Region}\label{sec:obsprob}

We first estimate the likelihood that JWST has already observed an overdense region similar to the simulated one within the JWST targeted fields. We compare the halo number density for the largest mass found in our simulation box with the effective volume probed by the sum of multiple survey areas at $z\sim30$. We consider the areas probed by the Cosmic Evolution Early Release Science Survey \citep[CEERS;][]{Finkelstein2025} in the Extended Groth Strip \citep[EGS;][]{Groth1994}, the JWST Advanced Deep Extragalactic Survey \citep[JADES;][]{Eisenstein2023} in the Great Observatory Origins Deep Survey \citep[GOODS;][]{Dickinson2003} North and South, the Public Release IMaging for Extragalactic Research \citep[PRIMER;][]{Donnan2024} in the Ultra Deep Survey \citep[UDS;][]{Lawrence2007}, and COSMOS-Web \citep{Casey2023} in the the Cosmic Evolution Survey \citep[COSMOS;][]{Scoville2007} fields.

CEERS probed 94 arcmin$^2$, JADES 220 arcmin$^2$, PRIMER 234 arcmin$^2$, and COSMOS-Web 1944 arcmin$^2$. The sum of these areas, $\sim$2500 arcmin$^2$, corresponds to a total observed volume of $\sim2.5\times10^{6}$ Mpc$^3$ comoving at $z\sim30.4$. This is the approximate volume required to detect at least one halo with mass of the order of the most massive halo ($\sim10^8$ M$_\odot$) in the simulation box, given the predicted number density of $\sim10^{-6}$ Mpc$^{-3}$ comoving (see Section~\ref{sec:biased}). It is, therefore, plausible to assume that existing JWST surveys have probed at least one such biased region around a similarly massive halo to the one simulated here.

\subsection{Cadence of Transient Events}\label{sec:eventprob}

From the simulations, we derive the number of PISNe events in post-processing based on the star formation rate density (SFRD), instead of directly counting the stellar particles formed throughout the simulation. We choose this approach, because each Pop III stellar particle represents an entire population, containing multiple massive stars that produce PISNe and core-collapse SNe at different times before the star particle's lifetime ends, where the total effect of all SN explosions from the population is complex and not fully resolved. To more robustly determine the number of PISNe across cosmic history, we measure the SFRD from the simulation and calculate the number of PISN events $N_{\rm PISN}$ per observed time per solid angle as \citep{Venditti2024}: 

\begin{equation}\label{eq:dndt}
    \frac{dN_{\rm PISN}}{dt_{\rm obs}d\Omega} = \frac{\overline{N}_{\rm PISN}}{M_{\rm III}}\int_{z_{\rm min}}^{z_{\rm max}} \frac{\Psi_{\rm III}(z)}{1+z}r^2(z) \frac{dr}{dz}dz\mbox{\ ,}
\end{equation}

where $\overline{N}_{\rm PISN}/M_{\rm III}$ is the average number of PISNe for a stellar population with total mass $M_{\rm III}$, $\Psi_{\rm III}(z)$ the SFRD, and $r(z)$ the comoving distance to redshift $z$. We estimate $\overline{N}_{\rm PISN}/M_{\rm III}$ as 
\begin{equation}\label{eq:pisnperstars}
     \frac{\overline{N}_{\rm PISN}}{M_{\rm III}} = \frac{\int^{260 ~\rm M_\odot}_{140 \rm~M_\odot}\phi(m)dm}{\int^{150 ~\rm M_\odot}_{1 \rm~M_\odot}m\phi(m)dm} 
\end{equation}
where $\phi(m)$ is the Pop~III IMF. We employ the IMF used in the simulation, given as \citep{Larson1998}
\begin{equation}
    \phi(m)\propto m^{-\alpha}\exp(-m^\beta_{\rm cut}/m^\beta)\mbox{\ ,}
\end{equation}
with $\alpha=0.17$, $\beta=2$, and $m^2_{\rm cut}=20$ M$_\odot^2$ for the mass range 1-150 M$_\odot$, resulting in $\overline{N}_{\rm PISN}/M_{\rm III}=9.4\times10^{-3}$ M$_\odot^{-1}$. However, the Pop~III IMF is uncertain \citep{Lazar2022}. E.g., considering a power-law IMF by setting $\beta=0$ and $\alpha=-0.17$ \citep{Jaacks2018,Venditti2024,Stacy2013}, results in $\overline{N}_{\rm PISN}/M_{\rm III}=1.2\times10^{-2}$ M$_\odot^{-1}$. As the detailed feedback recipe used in the simulation is not important here, given that we are examining the first star formation episodes, we vary the Pop~III IMF in our post-processing analysis.

\begin{figure}
\centering
\includegraphics[width=0.45\textwidth]{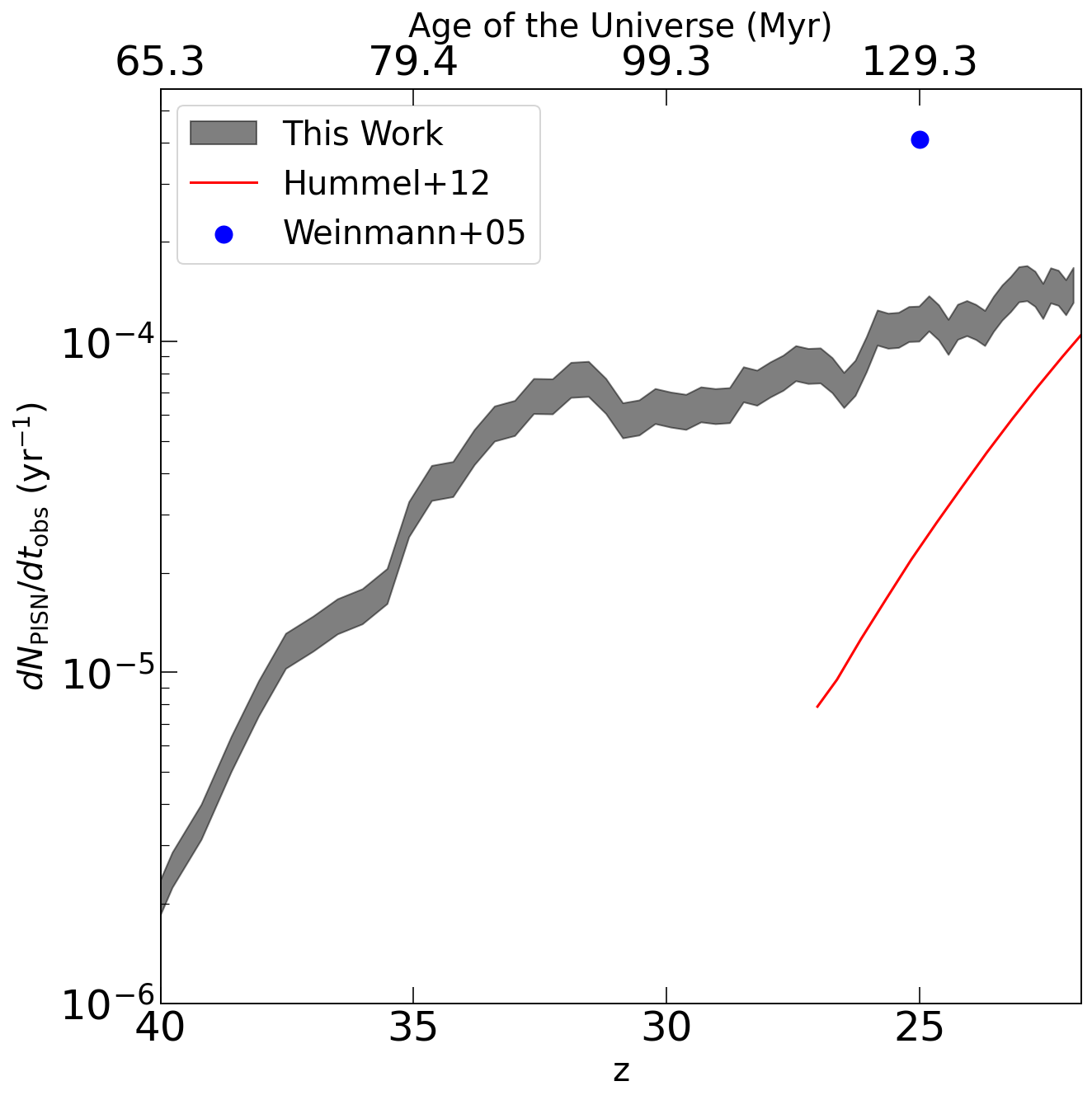}
    \caption{PISN rate across redshift inferred from our simulation. The shaded region reflects the range due to Pop~III IMF variation. We compare to predictions from previous work \citep{Weinmann2005,Hummel2012}, scaled to our simulation volume. \citet{Hummel2012} consider an average-density region of the Universe, resulting in rates that are lower than for the biased case considered here. \citet{Weinmann2005} predict a much higher rate, but with substantial uncertainties in their model assumptions. Across the whole redshift range spanned by the simulations $(99<z<22)$ and accounting for visibility time, cumulatively we expect $\sim10^{-1}$ events (see Section~\ref{sec:eventprob}). Thus, JWST could detect a PISN event as observations continue, at the tail-end of the probability distribution.}
    \label{fig:rate}
\end{figure}

We then measure the angular area probed by our 6 $h^{-1}$ cMpc box, corresponding to $6\times6$ $h^{-2}$ cMpc$^2$ surface area, across redshifts.  Using this angular size in the expression in Eq.~\ref{eq:dndt}, we derive $dN_{\rm PISN}/dt_{\rm obs}$, the number of PISN per unit observed time in the simulation. We show the resulting PISN rates across redshift in Fig.~\ref{fig:rate}, where the PISN rate increases with redshift following Pop III star formation. The shaded area reflects the variation resulting from the IMF choice. While the rates are low, $\sim10^{-4}$ events per observed year, the redshift regime we probe spans an order of $\sim10$ Myr in the restframe, representing a sufficiently long time interval for numerous PISNe to occur. We further compare our predictions to previous work \citep{Hummel2012,Weinmann2005}, scaled to our simulation volume. \citet{Hummel2012} predicts lower values at higher redshifts, but rates become comparable around $z\sim20$. This is plausible since \citet{Hummel2012} considered the PISN rate in the average Universe using a semi-analytic model focused on $10^9$ M$_\odot$ halos at $z\sim10$, and not on a biased region as we have done in this work. Given that predicted PISN rates are similar by $z\sim20$, the combination of an earlier onset of star formation and higher total rates in the overdense region may significantly improve chances of observing a PISN at very high redshift. At more intermediate redshifts, on the other hand, the possible boost by looking at overdense fields may not be as significant. Although \citet{Weinmann2005} predict a much higher PISN rate, they emphasize the substantial uncertainty in their predictions, arising from the assumptions on the IMF and resulting number of PISNe per stellar mass.

To determine whether JWST could have observed a PISN event during its initial period of operation, we set $z_{\rm max}=99$ and $z_{\rm min}=21.8$ in Eq.~\ref{eq:dndt}, the start and end redshifts of the simulation run. We thus probe the cumulative PISN rate per unit time in the biased region at extremely high redshifts. We find $dN_{\rm PISN}(99<z<21.8)/dt_{\rm obs}= 4.4\times10^{-3}-5.6\times10^{-3}$ yr$^{-1}$, depending on the IMF choice. If we further integrate across the optimistic visibility time of a PISN event with JWST at $z\gtrsim22$, $\Delta t_{\rm vis}\sim20$ years across redshift (see Section~\ref{sec:brightness} and Fig.~\ref{fig:magnitude}), we obtain the predicted number of PISNe in the biased region for $z\gtrsim22$, observable with JWST, as $N_{\rm PISN}(99<z<21.8)\sim0.1$ events.\footnote{If we limit the redshift range to $35<z<30$ to match the $z\sim32$ candidate \citep{Gandolfi2025}, we expect $\sim0.01$ events.} Therefore, JWST has a non-negligible chance to observe a PISN event at extremely high redshifts should it have observed a biased region, similar to the one simulated here. Indeed, it is highly likely that JWST has already observed at least one such region that could in principle host PISN events  (Section~\ref{sec:obsprob}). As JWST continues to operate, the chances of detecting such extremely early transients will increase, and the overdense fields will be ideal targets to hunt for PISNe at extremely high redshifts.


\subsection{Luminosity of Transient Events} \label{sec:brightness}

\begin{figure*}[!htb]
\centering
\includegraphics[width=0.8\textwidth]{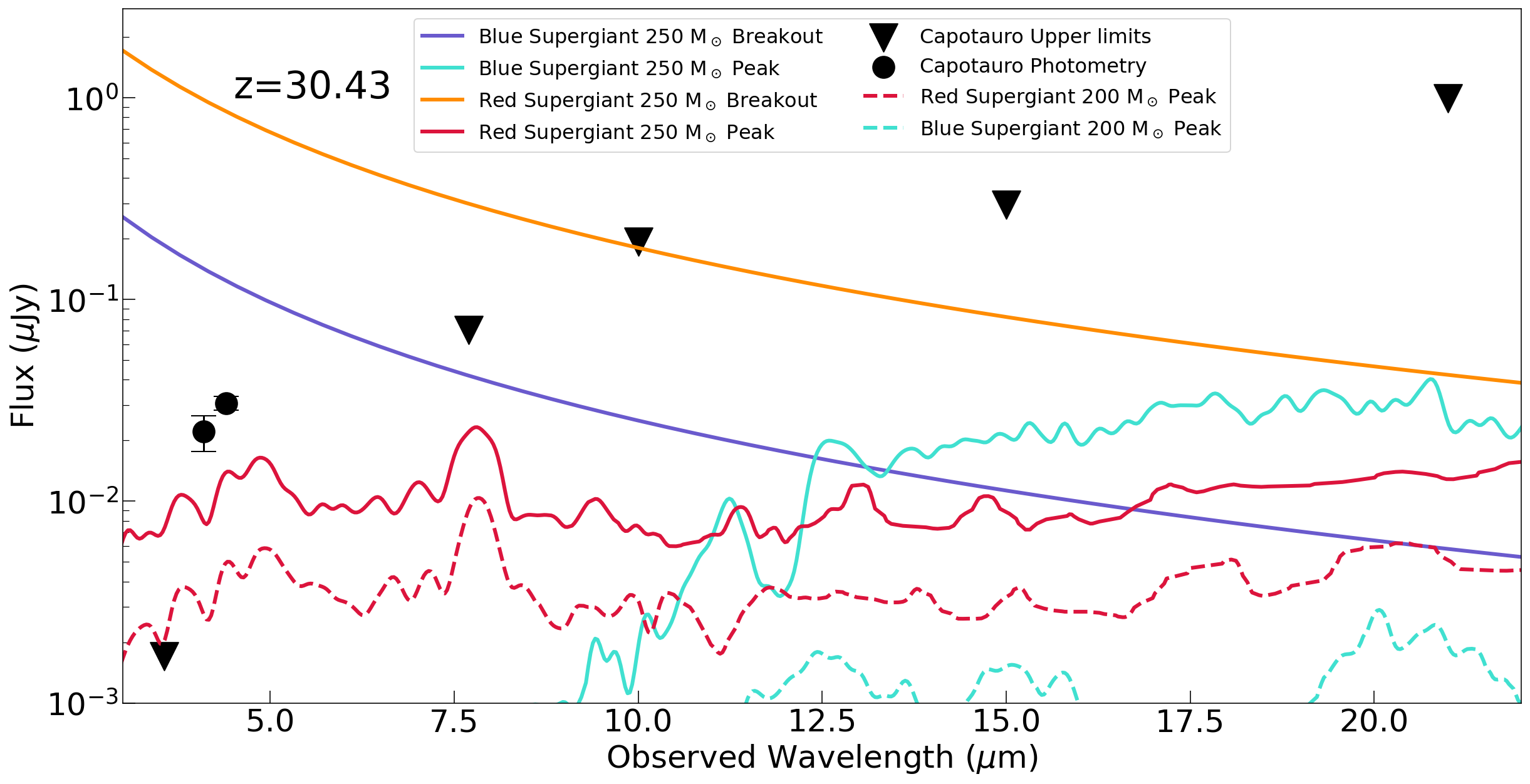}
    \caption{Model PISN spectra from a source at $z=30.4$. We reproduce spectra from \citet{Kasen2011}, corresponding to their B250 and B200 model, PISNe from a 250 M$_\odot$ and 200 M$_\odot$ metal-free blue supergiant, and the R250 and R200 model, PISNe from a 250 M$_\odot$ and 200 M$_\odot$ low-metallicity ($10^{-4}$ Z$_\odot$) red supergiant. We show the situation at breakout, when the explosion shock wave first reaches the surface, and at subsequent times, when the expanding and cooling ejecta produce bright emission. We display the spectra at peak luminosity following the breakout (17 days after breakout for the red and 383 days for the blue supergiant). We compare the model spectra with the observed photometry of Capotauro, proposed to be at $z\sim32$ \citep{Gandolfi2025}. The observed photometry and upper limits are comparable to the predicted PISN spectra for the most massive progenitors. Thus, if PISN events do occur at extremely high redshifts, they could be bright enough to be observable.}
    \label{fig:spectra}
\end{figure*}


Our results show that the simulated biased region could have been observed and that numerous PISN events are expected to occur in such a region. To be able to observe a possible PISN event, however, the explosion should be bright enough to be detected. To test the detectability of a PISN event, we take PISN model spectra from \citet{Kasen2011}, considering two cases: a PISN from a metal-free blue supergiant star (B250), and from a $10^{-4}$ Z$_\odot$ red supergiant (R250), to consider both metal-free and extremely metal-poor PISN progenitors. We consider the PISN to originate at $z=30.4$ and take the model spectra at breakout, when the shock wave first reaches the surface of the progenitor star's hydrogen envelope. The breakout phase is the brightest period of the PISN lightcurve, but it is also very brief, only lasting hours in the restframe. The breakout spectrum is modeled as a blackbody, with temperatures $6.3\times10^5$ K and $3.5\times10^5$ K for the blue and red supergiant cases, respectively \citep[according to][]{Kasen2011}. We further consider the model spectra after the breakout as the shock and hot ejecta from the explosion power the event. We compare the spectra at their peak luminosity after breakout, which occurs 17 days later for the red and 383 days later for the blue supergiant, and for the red supergiant case 100 and 300 days after the explosion. Fig.~\ref{fig:spectra} shows the PISN model spectra compared with photometric data of the Capotauro source, proposed to be at $z\sim32$ \citep{Gandolfi2025}. The observed photometry and upper limits are comparable to the model spectra of the most massive progenitors, so the PISN events for the most optimistic cases could be bright enough to be observable with JWST even at extremely high redshifts, even if Capotauro itself were not a $z\sim30$ PISN (see Sec.~\ref{sec:discussion}).

\begin{figure}
    \centering
    \includegraphics[width=0.5\textwidth]{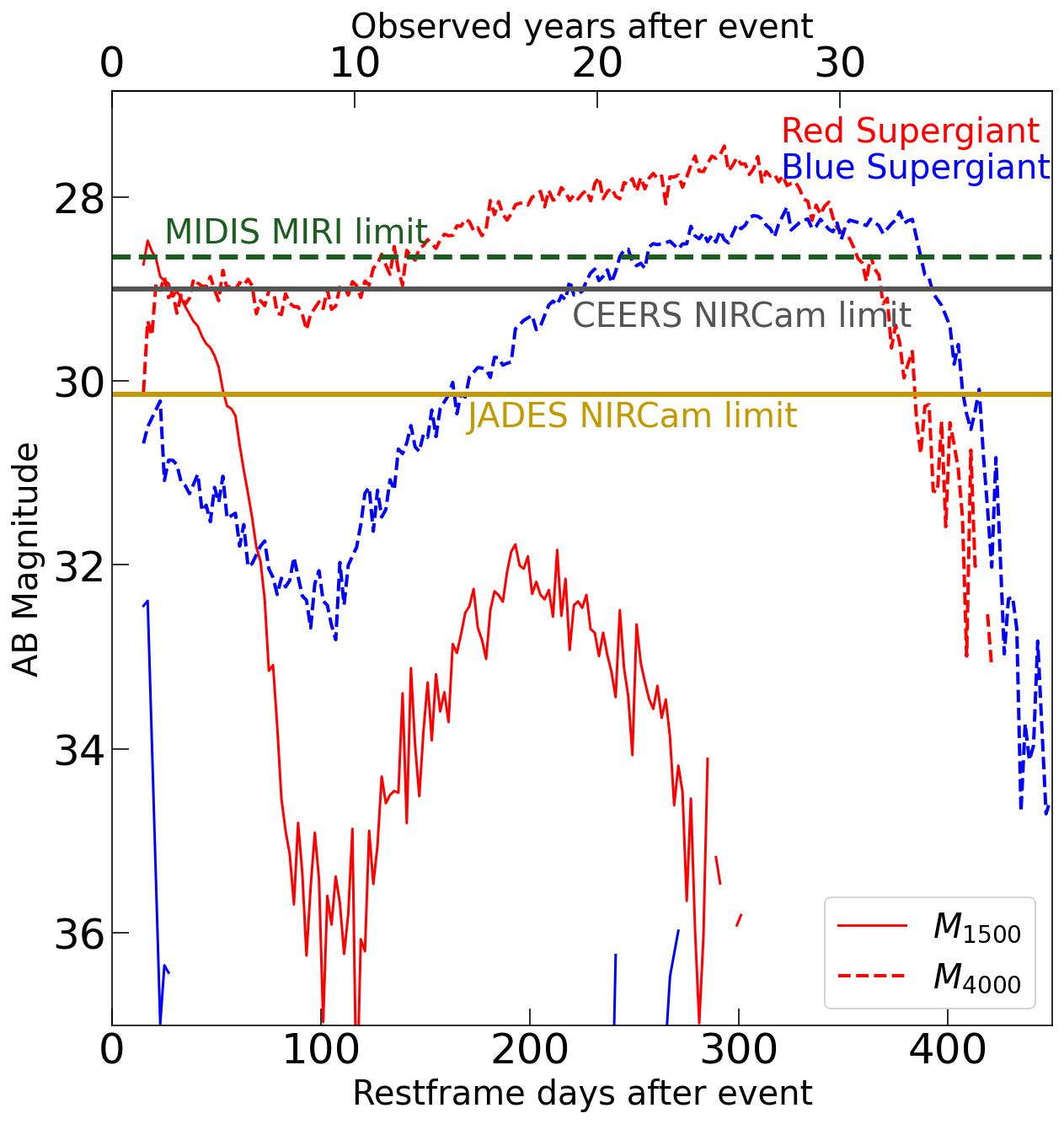}
    \caption{Brightness evolution of the model PISN spectra from \citet{Kasen2011} at $z=30.4$ across restframe days/observed years after the PISN explosion for the red and blue supergiant cases. We measure magnitudes around 1500 and 4000 \AA\ restframe, corresponding to wavelengths observable by JWST's NIRCam (1500 \AA) and MIRI (4000 \AA) instruments. We further show the magnitude limits of different JWST surveys. At their peak, PISN magnitudes are around 28-29, lasting for $\sim200$ days in the restframe and $\sim20$ years in the observed frame, which existing JWST surveys like CEERS, JADES, and the MIDIS have reached \citep{Finkelstein2025,Ostlin2025,Eisenstein2023}. Therefore, PISNe at extremely high redshifts may be observable with JWST, at least at peak brightness in their lightcurves.}
    \label{fig:magnitude}
\end{figure}

We further assess PISN observability by calculating their observed magnitudes at different wavelengths. We first estimate the average flux across wavelength/frequency as \citep{Papovich2001}
\begin{equation}
    \langle f_\nu\rangle = \frac{\int (t_\nu f_\nu/\nu)  d\nu}{\int(t_\nu/\nu) d\nu}\mbox{\ ,}
\end{equation}
where $f_\nu$ is the observed flux density, and the transmission function $t_\nu$ is a top-hat filter that is 1 around the chosen wavelength and 0 everywhere else. We then determine the AB magnitude from the average flux following $m_{\rm AB} = -2.5\log(\langle f_\nu\rangle)-48.60$. Fig.~\ref{fig:magnitude} shows the resulting magnitude evolution of the PISN for the red and blue supergiant models at $z=30.4$ across restframe days since the explosion. We measure the magnitudes at 1500 and 4000 \AA\ restframe with a filter width of 200 \AA, corresponding to $\sim4-12$ \micron\ in the observed frame, which the JWST Near Infrared Camera (NIRCam) can observe for the 1500 \AA\ emission, and the Mid Infrared Instrument (MIRI) for the 4000 \AA\ emission. We show the magnitude limits reached by the CEERS NIRCam \citep{Finkelstein2025}, JADES NIRCam \citep{Eisenstein2023}, and MIRI Deep Imaging Survey (MIDIS) MIRI \citep{Ostlin2025} programs, which lie below the PISN lightcurves at their peak around 28-29 magnitudes. 

\citet{Whalen2013} also predicted PISN magnitudes at $z\sim30$ allowing to be observed by NIRCam with peak magnitudes around 28-29, agreeing with our predictions. JWST could thus identify PISN events at least at their brightest phases in the lightcurve. While such peaks persist for $\sim200$ days in the restframe, at $z\sim30$ this corresponds to $\sim20$ years in the observed frame. On average, for $z\gtrsim22$, the visibility time is $\sim20$ years, long enough to be observed in principle. We note that we have only considered PISNe from the brightest 250 M$_\odot$ progenitors, thus choosing the most optimistic case to explore the limits of parameter space. In addition, PISNe from lower mass progenitors, while fainter at early times, may re-brighten at later times to $\sim29-30$ magnitudes to be observable even at $z\sim30$ \citep{Whalen2013}. Therefore, our optimistic estimates could apply to lower mass PISN progenitors as well.

\section{Detecting a Transient Event} \label{sec:discussion}

So far we have demonstrated that under current JWST programs, there is a non-negligible probability that a PISN event at extremely high redshifts $(z\sim30)$ could be observed. JWST surveys have covered large enough volumes to include extremely biased regions (Section~\ref{sec:biased}), and in such regions, numerous PISNe will have occurred over the local Hubble time (Section~\ref{sec:eventprob}). A subset of such PISN events is expected to be extremely bright, so that even at $z\sim30$, they should be observable (Section~\ref{sec:brightness}).

We note multiple caveats to be considered when interpreting our results. We assumed that to detect one biased region as we have simulated, JWST will need to observe all of the survey fields listed, CEERS, JADES, PRIMER, and COSMOS-Web. JWST will then have to continuously observe the biased region that may exist among the surveys to have a chance of detecting a transient event. However, JWST does not observe all survey fields at once, given its narrow field of view, and so the chances of observing a PISN event at extremely high redshifts may be lower than the estimate presented here. This optimistic limit could be reached if an overdense subregion within the JWST fields could be identified. Such a region would provide an ideal target to identify a PISN, since at extremely high redshifts, the expected PISN rate in overdense regions will be much higher than in the general Universe (Fig~\ref{fig:rate}).

Furthermore, our optimistic estimate depends on the star formation model, resolution, and the IMF we employed in our simulations. The simulation star formation subgrid model has been calibrated to high redshift observations and high-resolution simulations \citep{Jaacks2018,Jaacks2019,Liu2020,Bromm2013,Stacy2016,Klessen2023}, but the nature of metal-free Pop~III star formation and the assembly of the first galaxies is still uncertain \citep[e.g.][]{Finkelstein2024,Somerville2025,Jeong2025}. Fig.~\ref{fig:rate} further demonstrates that the choice of the Pop III IMF can affect the PISN rate predictions. Recent and near-future observations could shed more light on the nature of the first stars \citep{Visbal2025,Venditti2025,Zier2025,Nakajima2025,Cai2025}, which will provide more robust predictions for the first transients and their observability.

Lastly, we have not considered how a source may be robustly identified as a PISN \citep[e.g.,][]{Hartwig2018}. Above, we have argued for the detectability of PISN events at $z\sim30$, leaving open the challenge of their identification. Candidate sources at $z\sim30$ need to be confirmed, including Capotauro, proposed to be an extreme-redshift galaxy, and their nature is still under debate. These objects may be lower redshift galaxies, active galactic nuclei, or little red dot interlopers, and their interpretation as extreme high-redshift galaxies currently relies on limited photometric data. They may also be local brown dwarfs or even exoplanets rather than actual extreme-$z$ sources \citep{Gandolfi2025,Perez2025}. We note that if Capotauro indeed were a PISN, it might be compatible with a lower-redshift solution at $z\sim15$ \citep{Ferrara2026}. Thus, even if in the future an object is confirmed at extremely high redshifts, it may be difficult to conclusively characterize them as PISN events by photometry alone. Even though they are transient, due to cosmological time dilation we expect the decay in their light curves to be very slow in the observed frame, so that their nature may not be immediately clear. 

However, if an object were confirmed at $z\sim30$, a bright explosion would be one of the few sources that could produce the required luminosity to be detected even at such early times. 
We note that alternative explanations, such as vigorously accreting supermassive black holes (SMBHs), would be pushed to the limit as well. To form such massive SMBHs at $z\sim30$, non-standard formation channels such as primordial or direct-collapse black holes may be required \citep{Dayal2024,Zhang2025,Qin2025,Jeon2025,Jeon2025_lrd}. However, observing such SMBHs at $z\sim30$ will be difficult \citep{Pacucci2015,Natarajan2017,Barrow2018,Whalen2020}, and we refer the reader to the relevant literature for further discussion.

Absorption lines in the spectra (e.g., CaII) from metal-poor Pop III progenitors $(\sim10^{-4}$ Z$_\odot)$ could also help in PISN identification \citep{Kasen2011}. We expect the first stars to emerge at $z\sim30$, such that significant stellar mass and galaxy buildup does not yet occur so early on, and standard star formation is not expected to produce bright enough sources to be observed  \citep{Schauer2023}. If some sources are robustly confirmed to be at $z\sim30$, we show in this work that a PISN could be a possible explanation, being observable even at such high redshifts under (very) optimistic assumptions. This would evidently push the PISN scenario to the limit. However, it is not clear whether there are any other viable alternatives, without breaking standard cosmology or astrophysics. Even if Capotauro were not confirmed as a PISN at $z\sim32$, we predict that future JWST observations may identify multiple transient candidates at $z\gtrsim20$ in overdense regions, thus motivating future searches and follow-up observations of existing candidates.


\section{Conclusions} \label{sec:conclusions}

With the advent of new facilities like the JWST, observations have pushed farther and earlier in cosmic history, finding more distant objects than ever before. We have used cosmological simulations to test how far the frontier can be extended under current capabilities with JWST. Considering the PISN transients, extremely luminous events that could be observed out to $z\sim30$, we find that within current survey volumes and assuming continuous monitoring, JWST may detect such a transient event, providing a possible origin for candidate sources at extremely high redshifts \citep{Gandolfi2025,Perez2025}. 

Looking beyond, wide-field surveys have better capabilities to detect transient events. Roman and Euclid surveys will probe much larger areas and are expected to detect numerous transient events, in principle up to $z\sim12$ \citep[e.g.][]{Duffy2025,Wang2023_roman,Whalen2013}. However, the wide surveys do not probe as deep as JWST and will not detect objects as faint as JWST can. For the high-redshift frontier, such deep observations will be necessary \citep[e.g.,][]{Wang2012}. Thus, both deep and wide surveys will complement each other in our understanding of transient events and the early Universe. As we have shown, detection of high-$z$ transients may soon occur, at the tail end of the probability distribution, extending the observational frontier to even earlier times. If such a detection were made, it would provide a tantalizing glimpse into extreme astrophysics at the very beginning of star and galaxy formation.

\begin{acknowledgments}
We thank Daniel Kasen for providing the PISN spectral models, and Anthony Taylor, Ansh Gupta, Hollis Akins, and Tae Bong Jeong for helpful comments. The authors acknowledge the Texas Advanced Computing Center (TACC) for providing HPC resources under allocation AST23026.
\end{acknowledgments}

\bibliography{ms}{}
\bibliographystyle{aasjournal}



\end{document}